\documentclass[10pt,twoside,twocolumn,english,superscriptaddress, showkeys, aps]{revtex4-1}
\usepackage[T1]{fontenc}
\usepackage[latin9]{inputenc}
\usepackage{babel}
\usepackage{amsmath}
\usepackage{amssymb}
\usepackage{graphicx}
\usepackage[unicode=true]
 {hyperref}

\makeatletter

\providecommand{\tabularnewline}{\\}

\date{\today}

\makeatother

\begin{document}
\author{Yingchao Lu}
\email{yclu@lanl.gov} 
\affiliation{Los Alamos National Laboratory, Los Alamos, New Mexico, 87545, USA}
\affiliation{Department of Physics and Astronomy, Rice University, Houston, Texas 77005, USA}
\author{Hui Li}
\author{Kirk A. Flippo}
\author{Kwyntero Kelso}
\author{Andy Liao}
\author{Shengtai Li}
\affiliation{Los Alamos National Laboratory, Los Alamos, New Mexico, 87545, USA}
\author{Edison Liang}
\affiliation{Department of Physics and Astronomy, Rice University, Houston, Texas 77005, USA}
\title{MPRAD: A Monte Carlo and ray-tracing code for the proton radiography
in high-energy-density plasma experiments}
\begin{abstract}
Proton radiography is used in various high-energy-density (HED) plasma
experiments. In this paper, we describe a Monte Carlo and ray-tracing
simulation tool called MPRAD that can be used for modeling the deflection
of proton beams in arbitrary three dimensional electromagnetic fields,
as well as the diffusion of the proton beams by Coulomb scattering
and stopping power. The Coulomb scattering and stopping power models
in cold matter and fully ionized plasma are combined using interpolation.
We discuss the application of MPRAD in a few setups relevant to HED
plasma experiments where the plasma density can play a role in diffusing
the proton beams and affecting the prediction and interpretation of
the proton images. It is shown how the diffusion due to plasma density
can affect the resolution and dynamical range of the proton radiography.
\end{abstract}
\maketitle

\section{Introduction}

Proton radiography\citep{PRAD_D3He_1st_Li2006,PRAD_TNSA_1st_Zylstra2012}
is a diagnostic tool for time-resolved and spatial-resolved studies
of the electromagnetic field structures in inertial confinement fusion
(ICF) and high energy density (HED) plasmas. The information about
the morphology and strengths of electric and magnetic field is coded
in the deflection angle of the proton beams and alters the proton
flux after interaction with electromagnetic field. The proton flux
is then recorded on a detector. This kind of imaging technique has
been used to characterize the electromagnetic fields and carry out
measurements in a variety of experiments, including ICF implosion
capsules\citep{PRAD_direct_drive_Rygg2008,PRAD_direct_drive_Li2008,PRAD_indirect-drive_Li2009,PRAD_indirect-drive_Li2010,PRAD_indirect-drive_Li2012,PRAD_direct_drive_Seguin2012,PRAD_direct_drive_Igumenshchev2014,PRAD_direct_drive_Mackinnon2006,PRAD_direct_drive_Volpe2011,PRAD_direct_drive_Perez2009},
magnetic reconnection\citep{PRAD_reconn_Li2007,PRAD_reconn_Nilson2008,PRAD_reconn_Willingale2010,PRAD_reconn_Fiksel2014,PRAD_reconn_Rosenberg2015,PRAD_reconn_Rosenberg2015b},
self-generated magnetic fields through Biermann battery term\citep{PRAD_BB_Li2007,PRAD_BB_Cecchetti2009,PRAD_BB_Petrasso2009,PRAD_BB_Li2009,PRAD_BB_Li2013,PRAD_BB_Gao2019,PRAD_BB_Lu2019}
and plasma instabilities\citep{PRAD_RT_Manuel2012a,PRAD_RT_Gao2012,PRAD_Weibel_Kugland2013,PRAD_Weibel_Fox2013,PRAD_RT_Manuel2013,PRAD_RT_Gao2013,PRAD_Weibel_Park2015,PRAD_Weibel_Huntington2015},
non-ideal MHD effects\citep{PRAD_Nernst_Willingale2010,PRAD_nonideal_MHD_Lancia2014,PRAD_Nernst_Gao2015},
and laboratory dynamo experiments\citep{PRAD_dynamo_Tzeferacos2018,PRAD_dynamo_Tzeferacos2017}.

In ICF and HED experiments, two distinct types of proton sources have
been developed for high performance diagnostics. First, in a capsule
implosion, $\mathrm{DD}$(3MeV) and $\mathrm{D^{3}He}$(14.7MeV) protons
from fusion reaction driven by multiple laser beams. The protons leave
tracks in CR-39\citep{CR39_Sinenian2011,PRAD_D3He_1st_Li2006} which
is etched and scanned to get the absolute location and track characteristics
of each proton \citep{CR39_Sinenian2011}; Second, broadband proton
beams\citep{PRAD_TNSA_Flippo2010,PRAD_TNSA_1st_Zylstra2012} up to
60MeV are driven by ultra-intense ($>10^{18}\mathrm{W/cm^{2}}$) short
pulse laser beam through Target Normal Sheath Acceleration (TNSA)
mechanism, and the proton flux is recorded on the radiochromic film
pack with a sequence of proton energies. In general, the TNSA proton
backlighter offers better spatial and temporal resolution, while the
$\mathrm{D^{3}He}$ fusion-based techniques offers better spatial
uniformity and energy resolution.

The understanding of field structure from proton images is limited
by the fact that the images are a two dimensional mapping of the three
dimensional field distribution. The general mapping can be nonlinear,
degenerate and diffusive. Direct interpretation of the proton images
is achievable only under the assumptions of simple field geometries.
Some inverse-problem type of general techniques\citep{ALG_Kugland2012,ALG_Bott2017,ALG_Graziani2017}
have been developed to infer the integral quantities over the line
of sight, e.g., magnetic field perpendicular to line of sight ($\int d\boldsymbol{z}\times\boldsymbol{B}$)
or MHD current along the line of sight ($\int d\boldsymbol{z}\cdot\nabla\times\boldsymbol{B}$).
The comprehensive description of the inverse-problem type of techniques
for proton images of stochastic magnetic fields has been developed\citep{ALG_Bott2017}.
However, the caustic and diffusive regimes are still challenging for
inferring the fields. The primary focus of this paper to develop a
tool to understand the proton image in the diffusive regime, where
a ballistically propagating beam from the source is diffused by Coulomb
scattering and stopping power.

There are some general-purpose Monte Carlo toolkits, e.g. MCNP\citep{MCNP_Werner2018}
and GEANT4\citep{GEANT4}, and tools specifically for HED applications\citep{ALG_Volpe2012,ALG_Levy2015},
for forward modeling of proton radiography. MCNP and GEANT4 have the
features to model the energy lost and collisional scattering of protons
in cold matter, but corrections are needed for calculations of plasma
stopping power and scattering angle\citep{Manuel2012}. In this paper,
we take those corrections related to plasma into account and develop
a more accurate Monte Carlo and ray-tracing tool called MPRAD(multi-MeV
proton radiography) for forward modeling of proton radiography. We
make some approximations in the models for Coulomb scattering and
stopping power used in the code. Those are good approximations under
the condition that proton energy $E_{p}>1\mathrm{MeV}$, electron
temperature $\frac{kT_{e}}{545\mathrm{eV}}<E_{p}[\mathrm{MeV}]$,
density $\rho/A\ll10^{4}\mathrm{g/cc}$ ($A$ is the mass number of
the matter), and the transition layer between cold matter and fully
ionized plasma is thin compared to the rest of the system. This condition
covers the conditions of a range of ICF and HED experiments. MPRAD
is written in Python with MPI+OpenMPI parallelization among particles
or rays, using Cython and MPI4py package. Cython compilation for the
core part of the code is used to improve the performance. The output
data from plasma-dynamical modeling such as radiation-magnetohydrodynamics
or particle-in-cell(PIC) simulations can be imported into MPRAD. The
Python package from Yt-project\citep{YT_Turk2010} is used to read
the data from FLASH\citep{FLASH_Fryxell2000} simulations. Each MPI
process gets the whole data set of pre-calculated quantities. Each
thread solves the Monte Carlo transport for each particle (or ray
transport for each ray) independently. And the binned data (or final
quantities for the rays) is collected after each process and thread
completes the calculation for the targeted number of particles(or
rays). The process for making MPRAD an open source code is ongoing.
Our tool will be used for designing and analyzing the data for the
recent OMEGA experiments of magnetic field generation in shock-shear
type of targets\footnote{Lu et. al., 2019, in preparation}.

This paper is organized as follows. Sec \ref{sec:Feature} describes
the features of the MPRAD code, including the model for Coulomb scattering
and stopping power. In Sec \ref{sec:Benchmark} we perform benchmark
simulations for cold matter with MPRAD and compare the results with
MCNP simulations. Some applications and examples for the effect of
diffusion process on the proton radiography are discussed in Sec \ref{sec:applications}.
The summary is given in Sec \ref{sec:Conclusions}.

\section{Features of the code\label{sec:Feature}}

In MPRAD, we solve the relativistic equation of proton motion, i.e.
the evolution of position and velocity of the protons in the beam,
in electromagnetic field, similar to the features in the existing
tools\citep{GEANT4,MCNP_Werner2018,ALG_Levy2015,ALG_Kugland2012}.
In addition we implement the stopping power and Coulomb Scattering,
both in cold matter approximation and weakly interacting plasmas.
Pre-calculated quantities are used to speed up the large scale simulations.

\subsection{Stopping power\label{subsec:Stopping-power}}

In MPRAD code, we use the models for stopping power and energy-loss
straggling in the literatures\citep{Berger1993,Stopping_Bethe1930,Stopping_Li1993,Stopping_GERICKE2002,Stopping_strag_Bonderup1971}.
The relativistic effects of protons are taken into account to accurately
calculate the motion of non-relativistic to highly relativistic protons
with $\beta=v/c$, where $v$ is the velocity of the proton, and $c$
is the speed of light. The velocity for the proton beam is assumed
to be much higher than the electron thermal speed $v_{p}>v_{e,th}$,
which implies $\frac{E_{p}/m_{p}}{kT_{e}/m_{e}}>1$, i.e.

\begin{equation}
\frac{kT_{e}}{545\mathrm{eV}}<\frac{E_{p}}{1\mathrm{MeV}}\label{eq:temp}
\end{equation}
Under the $v_{p}>v_{e,th}$ assumption, we further assume that the
beam--plasma coupling strength\citep{Stopping_GERICKE2002} $\gamma_{c}$
is small, i.e.

\begin{equation}
\gamma_{c}=6.8\times10^{-3}\frac{(\frac{\rho}{1\mathrm{g/cc}}){}^{1/2}}{(\frac{E_{p}}{1\mathrm{MeV}}){}^{3/2}A^{1/2}}\ll1\label{eq:gamma}
\end{equation}
where $A$ is the mass number of the matter. For $E_{p}>1\mathrm{MeV}$
and $\rho/A\ll10^{4}\mathrm{g/cc},$ $\gamma_{c}$ is always much
less than unity, so that the beam--plasma coupling effect can be
neglected\citep{Stopping_GERICKE2002}.

For room temperature, we use the stopping power for cold matter. For
proton energy $E_{p}>1\mathrm{MeV}$, the stopping power in cold matter
can be written as\citep{Berger1993}
\begin{align}
\frac{d\langle E_{p}\rangle}{dx} & =-\frac{4\pi e^{4}n_{e}}{\beta^{2}m_{e}c^{2}}\bigg[f(\beta)+a\bigg]\nonumber \\
 & =-\frac{0.31\mathrm{MeV/cm}\times Z\frac{\rho}{1\mathrm{g/cc}}}{A\beta^{2}}\bigg[f(\beta)+a\bigg]
\end{align}
where $m_{e}$ is the mass of electron, $n_{e}$ is the total electron
number density(including both free electrons and bound electrons),
$\rho$ is the density of the matter, $A$ is the mass number of the
matter, and $Z$ is the charge number of the matter. The bracket $\langle E_{p}\rangle$
represents the average energy lost, and $x$ is the path length of
the proton. Due to the fact that the collision between the protons
and the particles in the matter is random, the energy lost follows
a distribution deviating from the average energy lost, which is described
as the straggling function as given in Eq(\ref{eq:straggling}). The
quantity $a$ is related to the material property and can be found
in stopping power database such as PSTAR \citep{PSTAR_Seltzer1993}
and SRIM\citep{SRIM_Ziegler2010}. The function $f(\beta)$ is 
\begin{equation}
f(\beta)=\ln[\beta^{2}/(1-\beta^{2})]
\end{equation}
 For mixture, compound or isotopes, i.e. different $A$'s and $Z$'s,
the stopping power is
\begin{equation}
\frac{d\langle E_{p}\rangle}{dx}=-\frac{0.31\mathrm{MeV/cm}\times Z_{1}\frac{\rho}{1\mathrm{g/cc}}}{A_{1}\beta^{2}}\bigg[f(\beta)+a_{\mathrm{CM}}\bigg]\label{eq:stopping_cold}
\end{equation}
where $f_{i}$ is the atomic number fraction of $i$th element, $A_{1}=\sum_{i}f_{i}A_{i}$
, $Z_{1}=\sum_{i}f_{i}Z_{i}$, and
\begin{equation}
a_{\mathrm{CM}}=\frac{\sum_{i}Z_{i}f_{i}a_{i}}{Z_{1}}\label{eq:a_CM}
\end{equation}
where the subscript ``CM'' denotes ``cold matter''.

For the calculations of plasma stopping power, only electron contribution
is considered, because the contribution of the plasma ions to stopping
power is neglectable due to the fact that $m_{i}/m_{e}=1836A\gg1$.
The expression of stopping power in plasma is simply the Bethe formula
under our assumptions in Eq(\ref{eq:temp}) and (\ref{eq:gamma})
\begin{align}
\frac{d\langle E_{p}\rangle}{dx} & =-\frac{4\pi e^{4}n_{e}}{\beta^{2}m_{e}c^{2}}\ln\bigg[1.123\sqrt{\frac{1}{2\pi}}\frac{m_{e}^{3/2}c^{2}\beta^{2}/(1-\beta^{2})}{\hbar e\sqrt{n_{e}}}\bigg]\nonumber \\
 & =-\frac{0.31\mathrm{MeV/cm}\times Z\frac{\rho}{1\mathrm{g/cc}}}{A\beta^{2}}\nonumber \\
 & \qquad\times\bigg[f(\beta)+10.2+0.5\ln A\nonumber \\
 & \qquad-0.5\ln Z-0.5\ln(\frac{\rho}{1\mathrm{g/cc}})\bigg]\label{eq:plasma-stop}
\end{align}
where $\hbar$ is the reduced Planck constant. Eq (\ref{eq:plasma-stop})
is consistent with the results from \citep{Stopping_Bethe1930,Stopping_Li1993,Stopping_GERICKE2002}.
For the plasma composed of multiple ion species
\begin{equation}
\frac{d\langle E_{p}\rangle}{dx}=-\frac{0.31\mathrm{MeV/cm}\times Z_{1}\frac{\rho}{1\mathrm{g/cc}}}{A_{1}\beta^{2}}\bigg[f(\beta)+a_{\mathrm{Plasma}}\bigg]
\end{equation}
where $A_{1}$ and $Z_{1}$ has the same definitions as in the cold
matter case but $f_{i}$'s are replaced by the number fractions of
ions. And 
\begin{align}
a_{\mathrm{plasma}} & =\frac{\sum_{i}f_{i}Z_{i}\big(10.2+0.5\ln A_{i}-0.5\ln Z_{i}\big)}{Z_{1}}\nonumber \\
 & \qquad-0.5\ln(\frac{\rho}{1\mathrm{g/cc}})\label{eq:a-plasma}
\end{align}

\begin{figure*}
\includegraphics[scale=0.3]{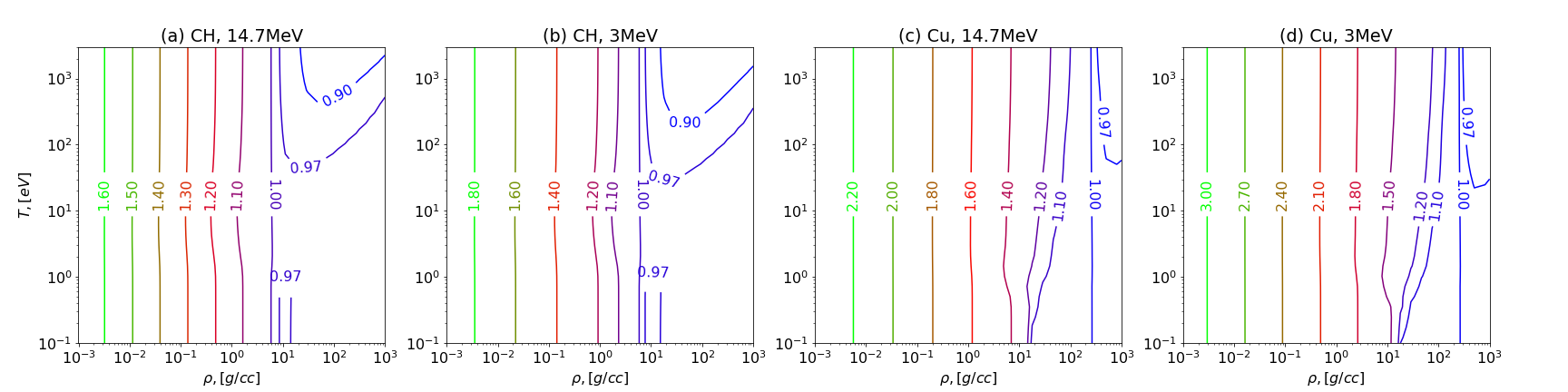}

\includegraphics[scale=0.3]{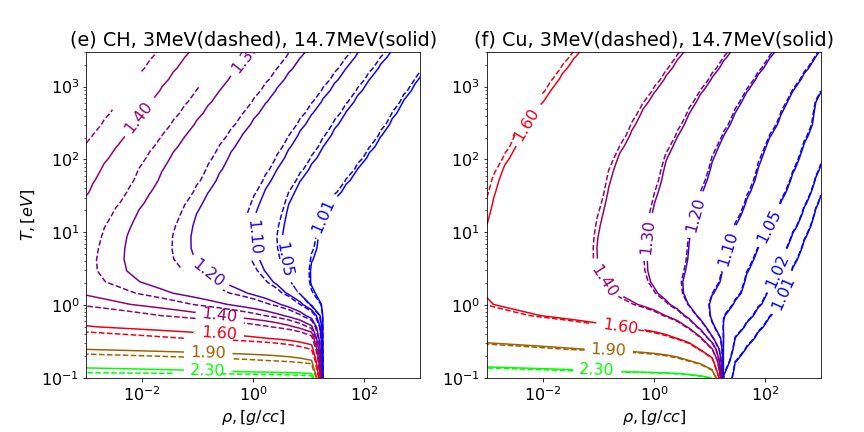}

\caption{Subfigures (a) to (d) are the contours of the ratio between the total
stopping power with the combined cold matter and plasma given by Eq(\ref{eq:stopping-combine})
and cold matter approximation given by Eq(\ref{eq:stopping_cold}),
for four different cases: (a) plastic(C:H=1), proton energy $E=14.7\mathrm{MeV}$,
(b) plastic(C:H=1), $E=3.0\mathrm{MeV}$, (c) copper, $E=14.7\mathrm{MeV}$,
(d) copper, $E=3\mathrm{MeV}$. Subfigure (e) and (f) are the contours
of the ratio between the scattering angle using the full characteristic
small scattering angle given by Eq(\ref{eq:small-angle}) and cold
matter approximation, for four different case (e) CH with proton energy
$E=14.7\mathrm{MeV}$ and $E=3\mathrm{MeV}$, (f) Cu with proton energy
$E=14.7\mathrm{MeV}$ and $E=3\mathrm{MeV}$. There is slight difference
between $E=14.7\mathrm{MeV}$ and $E=3\mathrm{MeV}$. The column density
is $\rho L=0.005\mathrm{g/cm^{2}}$ for (e) and (f). The horizontal
axes of all the subfigures are densities, and the vertical axes of
all the subfigures are electron temperatures.\label{fig:ratio-stopping}}

\end{figure*}

The difference between cold matter stopping power and plasma stopping
power is only in the expressions for $a_{\mathrm{CM}}$ and $a_{\mathrm{plasma}}$,
i.e. Eq(\ref{eq:a_CM}) and Eq(\ref{eq:a-plasma}), while other parts
of the two equations are identical. The typical values of $a_{\mathrm{CM}}$
or $a_{\mathrm{plasma}}$ are around 10. In a typical HED target system,
there are both cold matter and plasma. We use the ratio of Debye length
$\lambda_{D}$ to Fermi radius $a_{Z}$ for quantifying the partition
between cold matter and fully ionized plasma 
\begin{align}
\frac{\lambda_{D}}{a_{Z}} & =\frac{\sqrt{\frac{kT_{e}}{4\pi n_{e}^{f}e^{2}}}}{0.885a_{0}Z^{-1/3}}\nonumber \\
 & =\frac{0.5Z^{1/3}\sqrt{\frac{T}{1\mathrm{eV}}}}{\sqrt{\frac{n_{e}^{f}}{\mathrm{10^{23}/cc}}}}\label{eq:Debye-Borh-ratio}
\end{align}
where $a_{0}$ is the Bohr radius and $n_{e}^{f}$ is free electron
density, i.e. not including the bond electrons, which is different
from total electron density $n_{e}$, i.e. including both bond electrons
and free electrons. For matter composed of multiple elements, we use
logarithm averaged charge number $Z_{\mathrm{lg}}=\exp(\sum_{i}f_{i}\log Z_{i})$
in Eq(\ref{eq:Debye-Borh-ratio}). The total stopping power with combined
cold matter and plasma is
\begin{equation}
(\frac{d\langle E_{p}\rangle}{dx})_{\mathrm{total}}=-\frac{0.31\mathrm{MeV/cm}\times Z_{1}\frac{\rho}{1\mathrm{g/cc}}}{A_{1}\beta^{2}}\bigg[f(\beta)+a_{\mathrm{total}}\bigg]\label{eq:stopping-combine}
\end{equation}
where
\begin{equation}
a_{\mathrm{total}}=\frac{\lambda_{D}^{2}}{\lambda_{D}^{2}+a_{Z}^{2}}a_{\mathrm{CM}}+\frac{a_{Z}^{2}}{\lambda_{D}^{2}+a_{Z}^{2}}a_{\mathrm{plasma}}\label{eq:a-combination}
\end{equation}
The combination using Eq(\ref{eq:a-combination}) is a good approximation
if the transition layer between cold matter and plasma is thin compared
to the fully cold matter or fully plasma regions, i.e. most regions
in the modeling has $a_{Z}\ll\lambda_{D}$ or $a_{Z}\gg\lambda_{D}$.
If $a_{Z}\approx\lambda_{D}$ dominates, then one need more precise
combination model in the transition region. In Figure \ref{fig:ratio-stopping},
we plot the ratio between the total stopping power with combined cold
matter and plasma given by Eq(\ref{eq:stopping-combine}) and cold
matter approximation given by Eq(\ref{eq:stopping_cold}) for plastic(CH
with C:H=1) and copper, and for proton energy $14.7\mathrm{MeV}$
and $3\mathrm{MeV}$. We use the mean ionization state from PROPACEOS\footnote{PROPACEOS is available at \href{http://www.prism-cs.com}{http://www.prism-cs.com}}
equation of state table to calculate the free electron density. The
cold matter approximation is good for the matter near or above solid
density, i.e. $1\mathrm{g/cc}$ for CH and $8.9\mathrm{g/cc}$ for
Cu. For low densities, the correction from Eq(\ref{eq:plasma-stop})
has significant contribution to total stopping power, especially for
Cu. For high temperatures, i.e. $T>50\mathrm{eV}$, the correction
from Eq(\ref{eq:plasma-stop}) has larger contribution for CH than
for Cu.

The straggling function of the proton energy, i.e. the variation of
stopping power along the path of motion follows a Gaussian distribution\citep{Berger1993}
\begin{equation}
F(\Delta,s)=\frac{1}{\sqrt{2\pi}\Omega}\exp[-\frac{(\Delta-\Delta_{\mathrm{av}})^{2}}{2\Omega^{2}}]\label{eq:straggling}
\end{equation}
with a variance $\Omega^{2}$, and a mean value $\Delta_{\mathrm{av}}$
equal to the product of path length $s$ and the stopping power. The
expression for the variance is\citep{Berger1993,Stopping_strag_Bonderup1971}
\begin{align}
\Omega & =\sqrt{4\pi e^{4}n_{e}\frac{1-\beta^{2}/2}{1-\beta^{2}}s}\nonumber \\
 & =0.16\mathrm{MeV}\sqrt{\frac{n_{e}}{10^{23}/\mathrm{cc}}\frac{s}{1\mathrm{cm}}\frac{1-\beta^{2}/2}{1-\beta^{2}}}\label{eq:straggling-omega}
\end{align}
which only depends on the electron density $n_{e}$, the path length
$s$ and the normalized velocity $\beta$ of the proton.

In both Monte Carlo and ray-tracing calculations, three quantities
are pre-calculated, (1) the coefficient for stopping power $\frac{0.31\mathrm{MeV/cm}\times Z_{1}\frac{\rho}{1\mathrm{g/cc}}}{A_{1}}$,
(2) $a_{\mathrm{total}}$, (3) $4\pi e^{4}n_{e}$. In each time step
of the particle motion, in both Monte Carlo and ray-tracing, the energy
lost can be calculated from the pre-calculated quantities and the
current value of $\beta$ of the particle or ray, as given by Eq(\ref{eq:stopping-combine}).
In the Monte Carlo calculation, the straggling of proton energy is
sampled at each time step using Eq(\ref{eq:straggling}). In the ray-tracing
calculation, the total variable of proton energy is calculated by
the numerical integration
\begin{equation}
\Omega_{\mathrm{total}}^{2}=\sum_{\Delta s}4\pi e^{4}n_{e}\frac{1-\beta^{2}/2}{1-\beta^{2}}\Delta s
\end{equation}
where the value of $\beta$ is the current value for the particle
and $n_{e}$ is at the particle location.

\subsection{Coulomb Scattering}

The cross-section of single and multiple Coulomb scattering in thick
foil was studied in \citep{CScat_Bethe1953,CScat_Moliere1948}, which
has been used in GEANT4\citep{GEANT4} and MCNP\citep{MCNP_Werner2018}.
The cross-section for large angle scattering remains unchanged from
cold matter to plasma
\begin{equation}
Ns\sigma(\chi)\chi d\chi=2\chi_{c}^{2}\chi d\chi q(\chi)/\chi^{4}
\end{equation}
where $\sigma(\chi)\chi d\chi$ is the differential scattering cross
section into the angular interval $d\chi$ by each atom(or ion), $s$
is thickness of the material, $N$ is the number of scattering atoms(or
ions) per volume, $q$ is the ratio of actual to Rutherford scattering
and approaches unity for large angle scattering, and 
\begin{equation}
\chi_{c}^{2}=4\pi Nse^{4}Z(Z+1)/(pv)^{2}
\end{equation}
where $p$ is the proton momentum and $v$ the velocity of the proton
beam. The physical meaning of $\chi_{c}$, is that the total probability
of single scattering through an angle greater than $\chi_{c}$, is
exactly one. For mixture, compound or isotopes 
\begin{equation}
\chi_{c}^{2}=4\pi Nse^{4}(Z_{2}^{2}+Z_{1})/(pv)^{2}\label{eq:chic}
\end{equation}
where $Z_{2}=\sqrt{\sum_{i}Z_{i}^{2}f_{i}}$ and $Z_{1}$ is the same
as that in Eq(\ref{eq:a_CM}). The expression for the numerical value
of $\chi_{c}^{2}$ in terms of density $\rho$ is
\begin{equation}
\chi_{c}^{2}=1.8\times10^{-7}\times\frac{\rho s}{\mathrm{g/cm^{2}}}\frac{(Z_{2}^{2}+Z_{1})}{A}\frac{1-\beta^{2}}{\beta^{4}}
\end{equation}

For Coulomb scattering in plasmas, we replace the Fermi radius $a_{Z}$
of the atom with the Debye length $\lambda_{D}$ of the plasma in
the calculation of characteristic small scattering angle where $q(\chi)$
approaches zero. In general, for the regions with both cold matter
and plasma, the characteristic small scattering angle is 
\begin{equation}
\chi_{0}=\lambdabar\sqrt{\frac{1}{a_{Z}^{2}}+\frac{1}{\lambda_{D}^{2}}}\label{eq:small-angle}
\end{equation}
where $\lambdabar$ is the De Broglie wavelength of the proton. For
cold matter approximation, $\lambda_{D}/a_{Z}\to\infty$, Eq(\ref{eq:small-angle})
recovers the characteristic small scattering angle in cold matter,
which is identical to Eq(8) in Ref. \citep{CScat_Bethe1953}.

For thick target where many scattering events occur, $B_{c}>5$ for
the variable $B_{c}$ given by the following equations (we use $B_{c}$
instead of $B$ as in Ref. \citep{CScat_Bethe1953} to avoid confusion
with magnetic fields)
\begin{align}
B_{c}-\ln B_{c} & =b=\ln\frac{\chi_{c}^{2}}{1.167\chi_{a}^{2}}\label{eq:Bc}\\
\chi_{a}^{2} & =\chi_{0}^{2}\big(1.13+3.76(Z_{2}e^{2})^{2}/(\hbar v)^{2}\big)\label{eq:chia2}
\end{align}
where $3.76(Z_{2}e^{2})^{2}/(\hbar v)^{2}$ is the second order term
in the Born approximation. In $\lambda_{D}/a_{Z}\to\infty$ limit,
the expression for the numerical value of $e^{b}$ without second
order term in Born approximation is
\begin{equation}
e^{b}\approx\frac{6680\frac{\rho s}{\mathrm{g/cm^{2}}}(Z_{2}^{2}+Z_{1})}{\beta^{2}AZ_{\mathrm{lg}}^{2/3}}\label{eq:e-to-b-th}
\end{equation}
which is consistent with Eq(22) in Ref. \citep{CScat_Bethe1953}.
The distribution of the scattering angle $\theta$ is expanded in
a series of $B_{c}$. The tabulated numerical values of the distributions
are in Ref. \citep{CScat_Bethe1953}. We keep the first three terms,
i.e the Gaussian distribution (zeroth order term) with
\begin{equation}
\sigma_{\mathrm{Gauss}}=\chi_{c}(B_{c}/2)^{1/2}\label{eq:scat-sigma}
\end{equation}
and the terms in $\frac{1}{B_{c}}$ and $\frac{1}{B_{c}^{2}}$. The
distribution is closer to Gaussian distribution when $B_{c}$ becomes
larger. At each time step of proton motion, $B_{c}$ is calculated,
using the length step $\Delta s$ as the target thickness $s$ in
Eq(\ref{eq:chic}). In Monte Carlo calculation, if $B_{c}$ if small,
i.e $B_{c}<5$, we fallback to use the cross section for single scattering.
The method we implement for Coulomb scattering as a random process
has been used in other Monte Carlo codes such as MCNP\citep{MCNP_mokhov2002implementation}
and GEANT4\citep{GEANT4}. Two quantities in each cell are pre-calculated
before the Monte Carlo or ray-tracing calculations. (1) the coefficient
for large angle scattering cross section, i.e. $4\pi Ne^{4}(Z_{2}^{2}+Z_{1})$,
where $Z_{2}=\sqrt{\sum_{i}Z_{i}^{2}f_{i}}$, (2) the characteristic
small scattering angle $\chi_{0}$.

For ray tracing calculation, we use the numerical integration of the
R.H.S. of Eq(\ref{eq:chic}) and the ion+atom density weighted value
of $\ln\chi_{0}$
\begin{equation}
\ln\chi_{0}=\frac{\sum_{\Delta s}(N\ln\chi_{0})\big|_{\mathrm{local}}\Delta s}{\sum_{\Delta s}N\big|_{\mathrm{local}}\Delta s}
\end{equation}
which is an analog of Eq(16) in Ref. \citep{CScat_Bethe1953}. In
Figure \ref{fig:ratio-stopping}(e) and (f), we show the ratio between
the scattering angle using the full characteristic small scattering
angle given by Eq(\ref{eq:small-angle}) and using cold matter approximation.
The scattering angle is calculated using Eq(\ref{eq:scat-sigma}),
and $B_{c}$, $\chi_{c}$ are calculated by ray-tracing of proton
beam through a material with column density is $\rho L=0.005\mathrm{g/cm^{2}}$.
For low densities or high temperatures, the correction from finite
$\lambda_{D}$ has significant contribution to total scattering angle
as shown in the top left corner of Figure \ref{fig:ratio-stopping}(e)
and (f). The difference between $3\mathrm{MeV}$ and $14.7\mathrm{MeV}$
protons is more prominent in CH than in Cu, which can be explained
by the sensitivity of $b$ to the proton energy or proton velocity
given Eq(\ref{eq:Bc}) and Eq(\ref{eq:chia2}). For CH, $3.76(Z_{2}e^{2})^{2}/(\hbar v)^{2}$
is 0.6 for $E_{p}=3\mathrm{MeV}$ and 0.1 for $E_{p}=14.7\mathrm{MeV}$,
both less than 1.13, thus $e^{b}\sim\frac{1}{v^{2}(1.13+3.76(Z_{2}e^{2})^{2}/(\hbar v)^{2})}$
is sensitive to proton energy. For Cu, $3.76(Z_{2}e^{2})^{2}/(\hbar v)^{2}$
is 26 for $E_{p}=3\mathrm{MeV}$ and 5 for $E_{p}=14.7\mathrm{MeV}$,
both much larger than 1.13, thus $e^{b}\sim\frac{1}{v^{2}(1.13+3.76(Z_{2}e^{2})^{2}/(\hbar v)^{2})}\sim\frac{1}{v^{2}\times3.76(Z_{2}e^{2})^{2}/(\hbar v)^{2}}\sim\mathrm{constant}$.

\section{Benchmark against MCNP code for cold matter\label{sec:Benchmark}}

\begin{figure}
\includegraphics[scale=0.25]{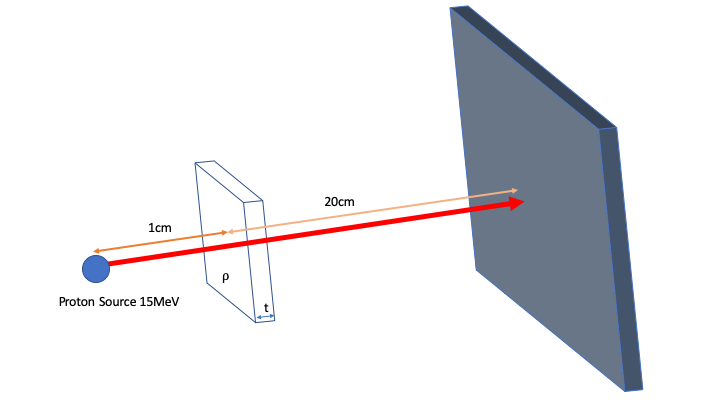}

\caption{Illustration of simulation setup for the benchmark of MPRAD against
MCNP and also for the example applications. The distance from the
proton source to the body center plane of the matter is 1cm. The density
of the matter is $\rho$ and the thickness is $t$. The detector plane
is parallel to the slab of matter and is 20cm away from the the body
center plane of the cold matter. The detector size is 2cm $\times$
2cm.\label{fig:setup}}
\end{figure}

Under cold matter approximation, we test the Monte Carlo calculation
in MPRAD code by the setup as shown in Fig \ref{fig:setup}. The mono-energetic($\Delta E_{p}=0$)
and collimated proton source with $E_{p}=15\mathrm{MeV}$ is placed
$1\mathrm{cm}$ from the slab of matter with given material and thickness.
We use $10^{6}$ particles in the simulations and the proton velocity
is perpendicular to the detector plane. The detector plane is $20\mathrm{cm}$
from the slab of matter, and the particles reaching the detector plane
are binned by spacial grid with $\Delta x=\Delta y=0.01\mathrm{cm}$
and energy grid with $\Delta E=0.05\mathrm{MeV}$. The simulation
with the same setup is also carried out using MCNP code\citep{MCNP_Werner2018}.

\begin{figure*}
\includegraphics[scale=0.4]{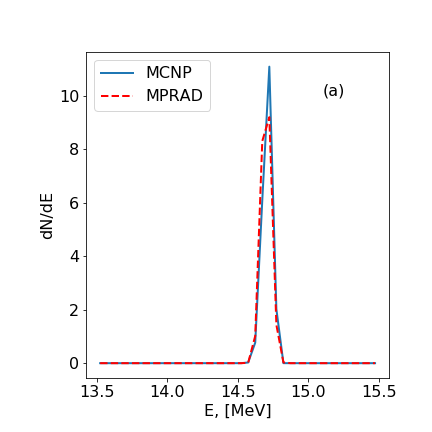}\includegraphics[scale=0.4]{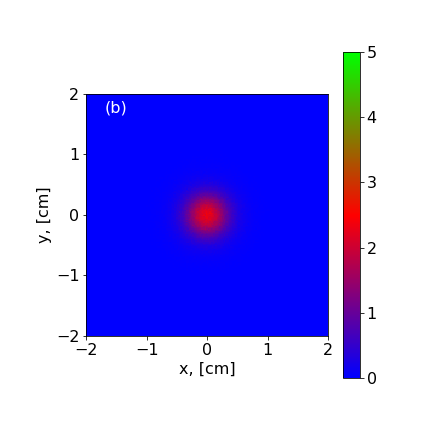}\includegraphics[scale=0.4]{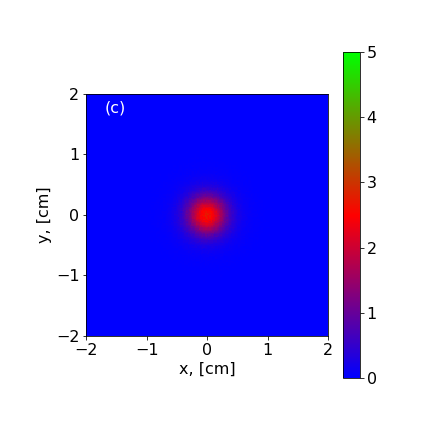}

\caption{Results of proton radiography simulation for a slab of $8.96\mathrm{g/cc}$
and $16\mathrm{\mu m}$ copper. (a) The energy grid is $\Delta E=0.05\mathrm{MeV}$.
The unit of the vertical axis is (number of particles)/$\mathrm{MeV}$
normalized by total particle number $N=10^{6}$. The red dash curve
is for MPRAD, the solid blue curve is for MCNP. (b) The image generated
using MPRAD, the spatial grid is $\Delta x=\Delta y=0.01\mathrm{cm}$.
(c) same as (b) but using MCNP. The unit of the color code is (number
of particles)/$\mathrm{cm^{2}}$ normalized by total particle number
$N=10^{6}$. \label{fig:benchmark}}
\end{figure*}

For all the test cases, both the spatially binned proton image and
the proton spectrum are consistent between MPRAD and MCNP. An example
is shown in Fig \ref{fig:benchmark}. The protons in the narrow beam
are isotropically scattered by colliding with the matter in the slab,
so a circular spot on the detector plane is produced as shown in Fig
\ref{fig:benchmark}(b) and (c). The protons lose energy and have
a finite width in the spectrum at the detector as shown in Fig \ref{fig:benchmark}(a),
because different protons have different path length in the matter
due to scattering. For a given composition of the slab material, different
density $\rho$ but same column density $\rho t$ produces similar
image and spectrum. Quantitative comparison between the results from
MPRAD and MCNP is shown in Table \ref{tab:Comparison-energy} and
Table \ref{tab:Comparison-width}. The slight difference between results
from MPRAD and MCNP is tolerant for typical proton radiography setup
in HED experiments, where the spectrum width of the source is a few
$\mathrm{keV}$ to $\mathrm{MeV}$\citep{Manuel2012,PRAD_TNSA_Flippo2010,PRAD_TNSA_1st_Zylstra2012}.

\begin{table*}
\caption{Comparison of average energy and energy variation from MPRAD and MCNP
for different material, density and thickness\label{tab:Comparison-energy}}

\begin{tabular}{|c|c|c|c|c|}
\hline 
Material, density, thickness & $\overline{E}$($\mathrm{MeV}$), MPRAD & $\overline{E}$($\mathrm{MeV}$), MCNP & $\sqrt{\overline{E^{2}}-\bar{E}^{2}}$($\mathrm{MeV}$), MPRAD & $\sqrt{\overline{E^{2}}-\bar{E}^{2}}$($\mathrm{MeV}$), MCNP\tabularnewline
\hline 
\hline 
Be, $1.85\mathrm{g/cc}$, $200\mathrm{\mu m}$ & 13.96 & 13.96 & 0.0546 & 0.0565\tabularnewline
\hline 
Be, $7.40\mathrm{g/cc}$, $50\mathrm{\mu m}$ & 13.96 & 13.96 & 0.0546 & 0.0565\tabularnewline
\hline 
Mg, $1.74\mathrm{g/cc}$, $16\mu m$ & 14.93 & 14.92 & 0.0157 & 0.0138\tabularnewline
\hline 
Mg, $6.96\mathrm{g/cc}$, $4\mu m$ & 14.93 & 14.92 & 0.0157 & 0.0138\tabularnewline
\hline 
Cu, $8.96\mathrm{g/cc}$, $16\mathrm{\mu m}$ & 14.70 & 14.71 & 0.0355 & 0.0350\tabularnewline
\hline 
Cu, $17.92\mathrm{g/cc}$, $8\mathrm{\mu m}$ & 14.70 & 14.71 & 0.0355 & 0.0350\tabularnewline
\hline 
\end{tabular}
\end{table*}

\begin{table*}
\caption{Comparison of weighted(by proton flux) average of $x^{2}+y^{2}$ from
MPRAD and MCNP for different material, density and thickness\label{tab:Comparison-width}}

\begin{tabular}{|c|c|c|}
\hline 
Material, density, thickness & $\sqrt{\overline{x^{2}+y^{2}}}$(cm), MPRAD & $\sqrt{\overline{x^{2}+y^{2}}}$(cm), MCNP\tabularnewline
\hline 
\hline 
Be, $1.85\mathrm{g/cc}$, $200\mathrm{\mu m}$ & 0.381 & 0.369\tabularnewline
\hline 
Be, $7.40\mathrm{g/cc}$, $50\mathrm{\mu m}$ & 0.381 & 0.369\tabularnewline
\hline 
Mg, $1.74\mathrm{g/cc}$, $16\mu m$ & 0.161 & 0.161\tabularnewline
\hline 
Mg, $6.96\mathrm{g/cc}$, $4\mu m$ & 0.161 & 0.161\tabularnewline
\hline 
Cu, $8.96\mathrm{g/cc}$, $16\mathrm{\mu m}$ & 0.535 & 0.536\tabularnewline
\hline 
Cu, $17.92\mathrm{g/cc}$, $8\mathrm{\mu m}$ & 0.535 & 0.536\tabularnewline
\hline 
\end{tabular}
\end{table*}

\section{Example applications\label{sec:applications}}

For small angle deflection, the deflection angle of protons by magnetic
field is \citep{ALG_Graziani2017}

\begin{align}
\alpha & =1.80\times10^{-2}\mathrm{rad}\times\big(\frac{E_{p}}{14.7\mathrm{MeV}}\big)^{^{-1/2}}\nonumber \\
 & \qquad\times\big(\frac{B}{10^{5}\mathrm{G}}\big)\times\big(\frac{l_{i}}{0.1\mathrm{cm}}\big)\label{eq:deflection-B}
\end{align}
where $E_{p}$ is the energy of proton, $B$ is the strength of magnetic
field, and $l_{i}$ is the longitudinal size of the interaction region.
From Eq(\ref{eq:scat-sigma}) and Eq(\ref{eq:deflection-B}) we can
calculate the ratio between the deflection angle by magnetic field
and the Coulomb scattering angle
\begin{align}
\frac{\alpha}{\sigma_{\mathrm{Gauss}}} & =\frac{2\sqrt{A}}{\sqrt{B_{c}(Z_{2}^{2}+Z_{1})}}\sqrt{\frac{E_{p}}{m_{p}c^{2}}}\frac{v_{A}}{c}\sqrt{\frac{l_{i}m_{p}c^{2}}{e^{2}}}\nonumber \\
 & =\frac{6\sqrt{A}}{\sqrt{B_{c}(Z_{2}^{2}+Z_{1})}}\big(\frac{E_{p}}{14.7\mathrm{MeV}}\big)^{^{1/2}}\nonumber \\
 & \qquad\times\text{\ensuremath{\frac{v_{A}}{2.8\times10^{4}\mathrm{cm/s}}}}\big(\frac{l_{i}}{0.1\mathrm{cm}}\big)\label{eq:angle-ratio}
\end{align}
where $v_{A}$ is the Alfvén speed, i.e. $v_{A}=\frac{B}{\sqrt{4\pi\rho}}$.

For the examples we show in this Section, we use the setup as shown
in Figure \ref{fig:setup}, with a magnetic field in the slab. We
carry out Monte Carlo runs with a flux rope of toroidal fields, a
flux rope of poloidal fields, and a turbulent field that satisfies
the power law energy spectrum. The $z$ axis is along the line of
sight, and the detector plane is $x-y$ plane. The interaction region
is filled with plastic(C:H=1). The thickness of the interaction region
is $l_{i}=1000\mathrm{\mu m}$ with uniform tunable density $\rho$
and fixed temperature $T_{e}=100\mathrm{eV}$. And the field is centered
at $(0,0,0)$. The source is at $(0,0,-1\mathrm{cm})$, mono-energetic
and collimated with $E_{p}=15\mathrm{MeV}$.

We use the same notation from Ref. \citep{ALG_Graziani2017} to define
the contrast field
\begin{equation}
\Lambda(\boldsymbol{x}_{\perp})=\frac{\Psi(\boldsymbol{x}_{\perp})-\psi_{0}}{\psi_{0}}
\end{equation}
where $\psi_{0}$ is the unperturbed proton flux, which is uniform
by assumption, $\Psi(\boldsymbol{x}_{\perp})$ is the perturbed proton
flux by both deflection and diffusion, and $\boldsymbol{x}_{\perp}$
is the position vector on the image plane. Eq(19) in Ref. \citep{ALG_Graziani2017}
gives the expression for the contrast field as a map of MHD current
\begin{equation}
\Lambda(\boldsymbol{x}_{\perp})=\frac{er_{i}(r_{s}-r_{i})}{r_{s}\sqrt{2m_{p}c^{2}E_{p}}}\hat{z}\cdot\int dz\ \nabla\times\boldsymbol{B}
\end{equation}
where $r_{s}$ is the distance between the interaction region and
the screen, $r_{i}$ is the distance between the source and the image
plate. For the parameters we use, we have
\begin{align}
\Lambda(\boldsymbol{x}_{\perp}) & =1.7\times10^{-6}\mathrm{G}^{-1}\times\big(\frac{E_{p}}{14.7\mathrm{MeV}}\big)^{^{-1/2}}\nonumber \\
 & \qquad\times\hat{z}\cdot\int dz\ \nabla\times\boldsymbol{B}\label{eq:contrast}
\end{align}
 We use the field strength that makes $|\Lambda(\boldsymbol{x}_{\perp})|<1$
to avoid caustics.

\begin{figure*}
\includegraphics[scale=0.37]{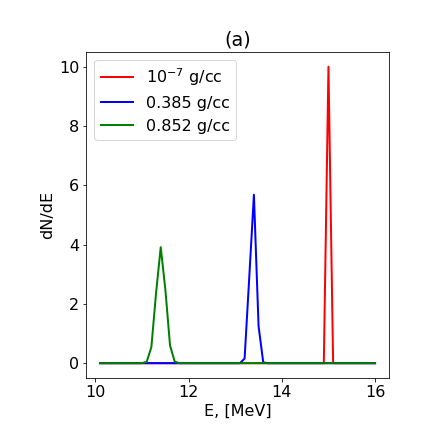}\includegraphics[scale=0.37]{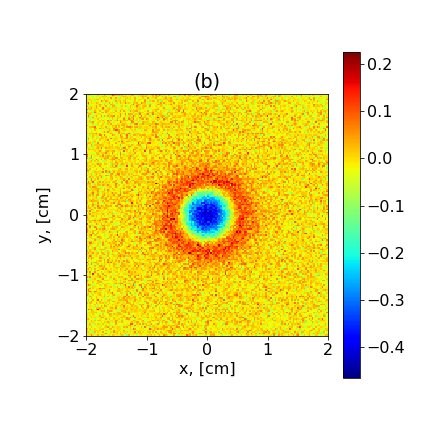}\includegraphics[scale=0.37]{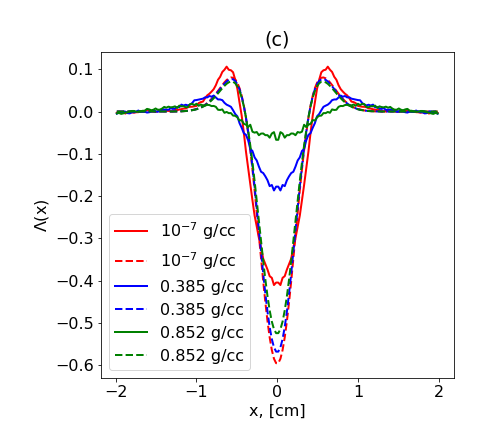}

\includegraphics[scale=0.37]{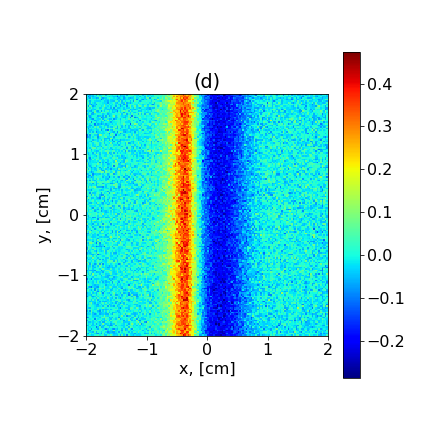}\includegraphics[scale=0.37]{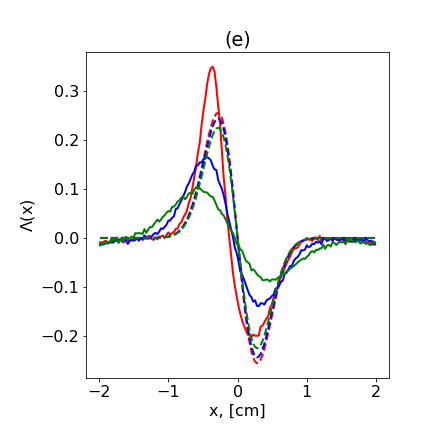}\includegraphics[scale=0.37]{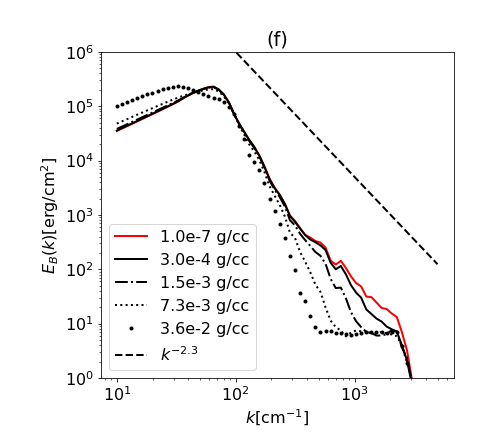}

\caption{The results for the example applications for $1000\mathrm{\mu m}$
thickness. The colors of the curves are consistent among all panels,
red for $\rho=10^{-7}\mathrm{g/cc}$, blue for $\rho=0.385\mathrm{g/cc}$,
green for $\rho=0.852\mathrm{g/cc}$, and black otherwise. Subfigure
(a) shows the spectrum of the protons in the detector plane for different
densities. Subfigure (b) is the contrast field of the proton image
for the test case for localized toroidal magnetic field, the feature
is coaxial as we can see from the symmetry of the field. A line-out
cross the center is shown in (c), for different densities. The dashed
lines are the theoretical value of contrast given by Eq(\ref{eq:contrast}),
and the solid lines are from the MPRAD simulations. The results for
the localized poloidal magnetic field is in (d) and (e). Subfigure
(f) is the inferred turbulence magnetic energy spectrum using Eq(\ref{eq:alg-inverse}),
for different densities. \label{fig:example-application}}
\end{figure*}

\subsection{Localized magnetic fields}

Toroidal magnetic fields have been observed and measured in some HED
experiments\citep{PRAD_BB_Cecchetti2009,PRAD_BB_Li2009,PRAD_BB_Petrasso2009,PRAD_D3He_1st_Li2006,PRAD_Nernst_Willingale2010,PRAD_nonideal_MHD_Lancia2014,PRAD_reconn_Fiksel2014,PRAD_reconn_Li2007,PRAD_reconn_Nilson2008,PRAD_reconn_Rosenberg2015,PRAD_reconn_Rosenberg2015b,PRAD_reconn_Willingale2010},
especially for the reconnection geometry. The typical geometry of
self-generated magnetic field in the plasma plume produced by single
laser spot is toroidal. We follow the expression for toroidal magnetic
field in literatures\citep{ALG_Graziani2017,ALG_Levy2015,ALG_Kugland2012},
which is the characteristic distribution of a localized toroidal field
\begin{equation}
\overrightarrow{B}=\frac{B_{0}}{a}\exp(-\frac{x^{2}+y^{2}+z^{2}}{a^{2}})(-y,x,0)\label{eq:B-toroidal}
\end{equation}
where we use $B_{0}=1\times10^{5}\mathrm{G}$, $a=200\mathrm{\mu m}$
for our numerical tests.

As shown in Figure \ref{fig:example-application}(a), the quasi-monoenergetic
proton beam has become a beam with a broad energy distribution as
it goes through the target region, and the mean energy becomes lower
than the source energy. In the analysis for real data from the experiments,
one also has to take the spectrum width of the source into account.
The diffusion can affect the interpretation of proton image. As the
density increases, the theoretical peak value of the contrast drops
as $E_{p}^{-1/2}$ given by Eq(\ref{eq:contrast}) if no diffusion
is considered. However, the peak value of the contrast in the simulation
drops faster than the theoretical value given by Eq(\ref{eq:contrast}),
as shown in Figure \ref{fig:example-application}(c). One deduces
smaller field or MHD current from the image for large density case.
For large densities, the variation level of proton number in each
pixel can potentially become comparable or even smaller than the poisson
noise for the CR-39 image, and the variation level of the proton flux
can become smaller than the sensitivity of radiochromic film.

A few HED experiments have generated and characterized poloidal magnetic
field, such as supersonic jets with mega-gauss self-generated magnetic
fields localized in the interaction region\citep{PRAD_BB_Gao2019,PRAD_BB_Lu2019}.
We follow the expression for poloidal magnetic field in literatures\citep{ALG_Levy2015,ALG_Kugland2012},
which is the characteristic distribution of a localized poloidal field

\begin{equation}
B_{y}=B_{0}\exp(-\frac{x^{2}+z^{2}}{a^{2}})\label{eq:B-poloidal}
\end{equation}
where we use $B_{0}=4\times10^{5}\mathrm{G}$, $a=200\mathrm{\mu m}$
for our numerical tests. The results are shown in Figure \ref{fig:example-application}(d)
and (e). Similar to the case for the toroidal magnetic fields, the
diffusion of the beam affect the final spectrum of the protons and
the peak value of proton flux contrast, and thus some care are needed
for interpreting the proton images.

\subsection{Power law energy spectrum in magnetic turbulence}

Magnetic turbulence and dynamo have been studied in HED experiments\citep{PRAD_dynamo_Tzeferacos2017,PRAD_dynamo_Tzeferacos2018}.
In turbulent magnetic fields, magnetic energy cascades to small scales,
and the magnetic energy spectrum follows a power law distribution.
The power law spectrum can be inferred using inverse-problem type
of technique\citep{ALG_Graziani2017,PRAD_dynamo_Tzeferacos2017,PRAD_dynamo_Tzeferacos2018}.
As an example for using MPRAD to study how diffusion affect the inferred
spectrum, we use a power law in a recently designed turbulent dynamo
experiment on the OMEGA-EP\citep{TMD_Liao2019}, where the magnetic
energy spectrum follows $E(k)\propto k^{-2.3}$. The method for generating
the magnetic field by random numbers for numerical tests is discussed
in Ref. \citep{ALG_Bott2017}, and the vector potential is multiplied
by $\exp(-\frac{x^{2}+y^{2}+z^{2}}{a^{2}})$ where $a=200\mathrm{\mu m}$
to get the localized field. We assume the RMS value of the magnetic
field in the $l_{i}^{3}$ box is $B_{\mathrm{rms}}=1\times10^{4}\mathrm{G}$,
the maximum field strength is $B_{\mathrm{max}}=1\times10^{5}\mathrm{G}$,
the same as the test problem for localized toroidal and poloidal magnetic
fields. We use the algorithm in Ref. \citep{ALG_Graziani2017} to
reconstruct the divergence free turbulence spectrum. For reconstruction,
Eq(52) in Ref. \citep{ALG_Graziani2017} gives the expression for
the inferred magnetic energy density
\begin{equation}
E_{B}(\frac{r_{s}}{r_{i}}\frac{2\pi}{L}|\boldsymbol{n}|)=\frac{2\pi}{e^{2}(r_{s}-r_{i})^{2}l_{i}L^{2}}\langle\hat{\Lambda}(\boldsymbol{n})\hat{\Lambda}(\boldsymbol{n})^{*}\rangle\label{eq:alg-inverse}
\end{equation}
where $l_{i}$ is the longitudinal size of the interaction region,
$L$ is the length and width of the image plate, $\hat{\Lambda}(\boldsymbol{n})$
is the discretized Fourier transform of $\Lambda(\boldsymbol{x}_{\perp})$,
and the average is over cells in $\boldsymbol{n}$-space for the discretized
Fourier transform.

As shown in Figure \ref{fig:example-application}(f), the diffusion
affects the cutoff length scale $\pi/k_{c}$ of the spectrum given
by the inversion algorithm. The results for $\rho=10^{-7}\mathrm{g/cc}$
shows little scattering and the spectrum around $k\sim10^{3}\mathrm{cm^{-1}}$
agrees being a the power law, and $k>3\times10^{3}\mathrm{cm^{-1}}$
is beyond the resolution limit. For densities from $\rho=3.0\times10^{-4}$
and above, there is a critical wavevector $k_{c}$ that the diffusion
affect damps the small scale feature for $k>k_{c}$ but retains the
large scale feature for $k<k_{c}$. For $\rho=7.3\times10^{-3}\mathrm{g/cc}$
and $\rho=3.6\times10^{-2}\mathrm{g/cc}$, the energy density at low
$k$, i.e. $k<50\mathrm{cm}^{-1}$ becomes higher than other densities.
The inverse of the cutoff scale $k_{c}$ is roughly the scattering
angle multiplied by $r_{i}$, thus

\begin{align}
k_{c} & \approx\frac{\pi}{10r_{i}\sigma_{\mathrm{Gauss}}}\nonumber \\
 & \approx20\mathrm{cm}^{-1}(\frac{\rho}{1\mathrm{g/cc}})^{-1/2}
\end{align}
where the factor $10$ in the denominator is an estimate of the scattering
angle in $\frac{1}{B_{c}}$ term. $B_{c}$ is roughly 6 for $\rho=1.5\times10^{-3}\mathrm{g/cc}$
so that $\frac{1}{B_{c}}$ term is not neglectable. The estimate for
$k_{c}$ is in good agreement with the results in Figure\ref{fig:example-application}(f).
In the analysis for real data from the experiments, one also has to
take the angular distribution of the source into account. The composition
and temperature can also affect $k_{c}$.

\section{Summary\label{sec:Conclusions}}

A simulation tool MPRAD is developed in this work, which extends the
capability of Monte Carlo calculations for proton radiography, especially
for the conditions where Coulomb scattering and stopping power are
not neglectable. The model for Coulomb scattering and stopping power
in fully ionized plasma and in cold matter are combined to improve
the accuracy of modeling, especially in the plasma region. Ray tracing
can be used as a quick way to study the effects of Coulomb scattering
and stopping power. Synthetic Monte Carlo radiograph by using the
imported data of fields, density, mass fraction and temperature distribution
from plasma-dynamical modeling can be useful for studying the interplay
between the effect from density and from electromagnetic fields. Such
kind of synthetic radiograph can aid optimizing the designs for experiments,
especially for the platforms where obstacle or high density plasma
is unavoidable.

TNSA protons with high energy are better to make the $\alpha/\sigma_{\mathrm{Gauss}}$
larger as given by Eq(\ref{eq:angle-ratio}), thus the deflection
by electromagnetic fields is more prominent than the diffusion of
the beam. The proton beam from fusion source can still be useful for
the magnetic field measurement, although the dynamical range is affected
by the diffusion. The signal on the proton image appears in a wide
range of energy band instead of a narrow band as the original beam,
which potentially gives us more information about the magnetic fields
for reconstructing field structure. However, obtaining the proton
images for different energy bands put challenges on the etching process
of CR-39\citep{CR39_Sinenian2011}. The Monte Carlo simulations with
scattering and energy lost included are needed for optimizing the
etching process.

\section{Acknowledgements}

YL is grateful to Alex Zylstra, Hong Sio, Andrew Birkel, Sky Sjue,
Mario Manuel, Don Lamb, Petros Tzeferacos and Matthew Baring for valuable
discussions. Research presented in this paper was supported by the
Laboratory Directed Research and Development(LDRD) program of Los
Alamos National Laboratory(LANL). The simulations were performed with
LANL Institutional Computing which is supported by the U.S. Department
of Energy National Nuclear Security Administration under Contract
No. 89233218CNA000001, and with the Extreme Science and Engineering
Discovery Environment (XSEDE), which is supported by National Science
Foundation(NSF) grant number ACI-1548562.

\bibliographystyle{apsrev4-1}
\bibliography{prad_sim}

\end{document}